\documentclass[aps,prl,twocolumn,floatfix,superscriptaddress]{revtex4}
\usepackage{graphicx}% Include figure files
\usepackage{dcolumn}% Align table columns on decimal point
\usepackage{bm}% bold math

\newcommand{\diffl}[2]{\frac{d #1}{d #2}}
\newcommand{\dfrac}[2]{\displaystyle\frac{#1}{#2}}

\newcommand{\kB}{k_{\text{B}}}
\newcommand{\thetaD}{\theta_{\text{D}}}
\newcommand{\thetaE}{\theta_{\text{E}}}
\newcommand{\rWS}{r_{\text{WS}}}

\begin{document}

%\preprint{Manuscript ??????}

\title{Equations of state for ruthenium and rhodium}

\date{July 29, 2019 %-- draft}%, IM\#974981}
   -- LLNL-JRNL-780237}

\author{Damian~C.~Swift}
\author{Thomas~Lockard}
\affiliation{%
   Lawrence Livermore National Laboratory,
   7000 East Avenue, Livermore, California 94551, USA
}
\author{Olivier~Heuz\'e}
\affiliation{%
   CEA/DAM-\^Ile de France,
   Bruy\`eres-le-Ch\^atel, F-91297 Arpajon Cedex, France
}
\author{Mungo~Frost}
\author{Siegfried~Glenzer}
\affiliation{%
   SLAC National Accelerator Laboratory,
   Menlo Park, California 94025, USA
}
\author{Kenneth~J.~McClellan}
\affiliation{%
   Los Alamos National Laboratory,
   MS~G770, Los Alamos, New Mexico 87545, USA
}
\author{Sebastien~Hamel}
\author{John~E.~Klepeis}
\author{Lorin~X.~Benedict}
\author{Philip~A.~Sterne}
\affiliation{%
   Lawrence Livermore National Laboratory,
   7000 East Avenue, Livermore, California 94551, USA
}
\author{Graeme~J.~Ackland}
\affiliation{%
   Department of Physics, University of Edinburgh,
   Edinburgh, EH9~3JZ, Scotland, UK
}

\begin{abstract}
Ru and Rh are interesting cases for comparing equations of state (EOS),
because most general-purpose EOS are semi-empirical, relying heavily
on shock data, and none has been reported for Ru.
EOS were calculated for both elements using all-electron atom-in-jellium theory,
and cold compression curves were calculated for the common crystal types
using the multi-ion pseudopotential approach.
Previous EOS constructed for these elements used Thomas-Fermi (TF) theory
for the electronic behavior at high temperatures, which neglects 
electronic shell structure;
the atom-in-jellium EOS exhibited pronounced features from the excitation
of successive electron shells.
Otherwise, the EOS matched surprisingly well, especially considering the
lack of experimental data for Ru.
The TF-based EOS for Ru may however be inaccurate in the 
multi-terapascal range needed for some high energy density experiments.
The multi-ion calculations predicted that the hexagonal close-packed phase
of Ru remains stable to at least 2.5\,TPa and possibly 10\,TPa,
and that its $c/a$ should gradually increase to the ideal value.
A method was devised to estimate the variation in Debye temperature $\thetaD$
from the cold curve, and thus estimate the ion-thermal EOS
without requiring relatively expensive dynamical force calculations, 
in a form convenient for adjusting EOS or phase boundaries.
$\thetaD$  estimated in this way was similar to the result from 
atom-in-jellium calculations.
We also predict the high-pressure melt loci of both elements.
\end{abstract}

%\pacs{07.35.+k, 52.38.Mf, 47.40.Nm, 79.20.Ds}

\maketitle

\section{Introduction}
Although they are relatively uncommon metals, Rh and Ru have several notable
technological applications involving dynamic loading and elevated states
of compression and heating.

Both Rh and Ru have important uses in strengthening alloys and
as coatings resistant to corrosion.
Their isotopes include some of the principal fission products of the actinides, 
and significant amounts build up as ingrowth in nuclear fuels and heat sources.
For this reason, Rh and Ru are relevant in the reprocessing of nuclear waste,
and their alloys are potential components of advanced nuclear fuels.
There are situations where these materials may be subjected to compression, 
heating, and ablation,
including aerospace applications of refractory alloys containing these elements,
and accident scenarios involving future space fuels.

Both Ru and Rh have been proposed as a key component in experiments
at the National Ignition Facility \cite{Prisbrey20xx}.
In these experiments, 
they are subjected to shocks of up to $\sim$10\,TPa,
and their subsequent expansion, along with that of lower-density components, 
is used to induce ramp loading in a sample to be studied \cite{Prisbrey2012}.
An experiment using Rh has been performed, though not completely analyzed
\cite{Park2019}.
These pressures are far above any calibration of the equation of state (EOS) 
used in the design and interpretation of these rather expensive experiments,
and for that matter well above almost all experimental data for any material.
Relative shock Hugoniot experiments have been performed to higher pressures
using shocks driven by nuclear explosions \cite{nim}, though with quite large
uncertainties and without an accurate calibration of the impedance-matching
standard materials.
Absolute Hugoniot experiments have recently been demonstrated in this
pressure range and above using laser-driven shocks
\cite{Doeppner2018,Swift_gbar_2018}, but involve x-ray radiography, 
and are not yet feasible on samples of such high atomic number.
This restriction is a motivation to evaluate the quality of existing EOS
in this regime, and to improve them if possible.

In many experiments using ablation to induce loading,
a variety of elements are employed as shields against unwanted x-ray heating
from the ablation plasma, chosen on the basis of the expected source spectrum
and deposition profile in the sample,
for instance in x-ray diffraction experiments \cite{McNaney2019}.
Although the accuracy EOS is often less important than for other components
of the target, it may be crucial to predict correctly whether the shield
melts, in order to identify the diffraction lines from the sample.
For this reason, it is desirable to construct the melt locus of possible
shield materials to high pressures.

Experiments have been performed on the intermetallic compound RuAl
using laser ablation to induce a shock, and measuring the flow stress and
spall strength. A model of Ru in the ablation plasma is needed
to interpret these measurements \cite{Swift_RuAl}.

There is a relative paucity of high pressure measurements on these elements.
Very unusually for a non-toxic engineering metal element,
we have found no shock data at all for Ru.
The comparison between EOS for these elements
is thus an interesting test of the performance of theoretical techniques 
with limited experimental data to constrain models.
We can make testable predictions of the high pressure properties
of Ru in particular that are not informed by experimental measurements.

Here we construct wide-ranging EOS for Ru and Rh 
using average atom techniques
most applicable in the fluid-plasma region, and assess their
accuracy against published experimental measurements and also against
multi-ion methods that perform better in the solid.
We also use multi-ion electronic structure calculations to construct
narrower-range EOS that apply better in the solid,
and use these results to extend previous predictions of the stability field 
of the ambient phase to much higher pressures.

\section{Previous equation of state studies}
Currently, all accurate, general-purpose EOS are semi-empirical.
The most usual thermodynamic approach is based on evaluating the
Helmholtz free energy $f$ as a function of mass density $\rho$ and
temperature $T$, decomposed into a cold compression curve $f_c(\rho)$
along with thermal contributions for the ions $f_i(\rho,T)$ and
electrons $f_e(\rho,T)$.
The pressure $p$ and specific internal energy $e$ are obtained by differentiation,
and it is the resulting tabulations $\{p,e\}(\rho,T)$ that constitute the
EOS used in practice for studies of impacts and other hydrodynamic situations.
Theoretical methods typically have difficulty in adequately reproducing the
observed mass density at standard temperature and pressure (STP),
and common practice has been to use the observed value, along with the
compressibility and shock data, to calibrate algebraic functions
for $f_c$ by taking simplified forms for $f_i$ in particular
a Gr\"uneisen form with a slowly-varying, analytic $\Gamma(\rho)$.
Observations and predictions of phase changes, informed by electronic
structure theory, have been used to make multiphase constructions
primarily of $f_c$ and $f_i$.
For $f_e$, Thomas-Fermi (TF) theory \cite{tf} has been used most commonly,
although some EOS have been constructed using treatments that, unlike TF, 
take account of electron shells.

For Rh, measurements of states along the principal shock Hugoniot have been
reported for pressures up to $\sim$0.2\,TPa \cite{McQueen1970}; 
for Ru there are none.
Room-temperature compression measurements have been reported for
both elements, but up to only 64 and 56\,GPa for Rh and Ru respectively
\cite{Clenenden1964,Yusenko2019,Tkacz1998,Cynn2002}.
Data from presses have not been used as extensively as shock data
for constructing EOS, because presses require a pressure calibrant
whose uncertainty introduces another source of error whereas
shock data can be absolute (and are the ultimate source of calibration
for press data), and also because applications of EOS largely involve
shock loading, so a direct calibration bypasses inaccuracies in treating
the region between the ambient isotherm and the principal Hugoniot.

The widely-used {\sc sesame} EOS library \cite{sesame}
includes a model for Rh \cite{Johnson1994}, constructed by the usual
semi-empirical approach using measurements of the principal shock Hugoniot
to deduce the cold compression curve, and blending into
a TF treatment at high temperatures.
The {\sc leos} library \cite{leos} has a model for Rh constructed similarly.
It also includes a model for Ru constructed using the 
`quotidian EOS' (QEOS) procedure \cite{qeos} 
again employing TF theory at high temperatures,
but using the measured mass density and sound speed at STP
in the absence of shock data.
No {\sc sesame} EOS has been constructed for Ru.

Theoretical studies of Ru and Rh have been included among extensive research
on the transition elements \cite{Tripathi1988,Cazorla2008,Cazorla2011}, 
although we have found no comparisons made
with existing EOS models, and a relative lack of work on the thermal EOS.
In most cases, no EOS was constructed, 
and results were insufficient for constructing EOS
into the fluid and plasma regimes.
For Rh, a mean field approach has been used to construct the EOS for the solid
\cite{Kumar2016},
though the resulting EOS was a poor match to mechanical properties 
such as observed shock Hugoniot states.
These calculations suggested that electron-thermal energy is unusually large
at $\sim$30\%\ of the ion-thermal energy above 800\,K.
The derivation of the electron-thermal energy was not described in detail,
but was apparently a free electron gas,
which is unlikely to be accurate over a wide range of states.
Studies using tight-binding theory \cite{Cazorla2008}
reproduced the observed crystal structure of Ru and Rh to the limits
of published data ($\sim$50\,GPa), and predicted that their ambient
structures would persist to at least 0.4 and 0.5\,TPa respectively.

Further work is clearly needed to test and constrain EOS into the 
high energy density regime.
A challenge is that the most rigorous techniques available are 
both computationally expensive and not strictly
valid over the full range from STP to states of high compression and heating,
and subsequent expansion.
Ru and Rh have atomic numbers 44 and 45 respectively, and so the 
electron-thermal EOS must dominate over the ion-thermal under shock loading
or heating sufficient to start ionizing the weakly-bound outer electrons.
The most rigorous approach to calculating electronic states is quantum
Monte Carlo, such as path integral Monte Carlo (PIMC) \cite{pimc}
which accounts for the antisymmetry of the electron wavefunctions under
particle exchange.
However, PIMC calculations have not been reported yet for such high atomic
numbers.
In PIMC, the ion-thermal energy is approximated by the high temperature limit
of an ideal monatomic gas,
$\frac 32 \kB T$ per atom, which is not appropriate in condensed phases.
At lower temperatures, quantum molecular dynamics (QMD) has been extensively
used, in which the effects of exchange and correlation on the wavefunctions
are estimated using Kohn-Sham density functional theory (DFT),
the inner electrons of each atom are subsumed into a pseudopotential
rather than treated explicitly, and ion trajectories are simulated from
Hellmann-Feynman forces on the ions \cite{qmd}.
The QMD treatment of the electrons is less rigorous than in PIMC and
the converse for the ions.
However, the ion trajectories are nevertheless classical, and do not
account for the quantum mechanical ion effects of 
zero-point motion and quenching of vibrational modes at low temperature.
Both of these effects can be accounted for by using Hellmann-Feynman forces
in DFT to calculate phonon modes for solid phases only \cite{Swift2001},
although, when constructing an EOS, integrating over 
the complete phonon spectrum means that the detailed contribution from
each mode is lost, and it is often adequate to describe the
ion-thermal free energy in terms of a Debye model \cite{Debye1912}
in which the Debye temperature $\thetaD$ depends on compression.
Formally, since the ion-thermal heat capacity must vary between zero at
low temperatures and $3\kB$ per atom when all the phonon modes
are fully excited, it is always possible to express the ion-thermal free energy
using a Debye model which depends on both density and temperature
$\thetaD(\rho,T)$.
In fact, as the bulk modulus of matter can also be regarded as an integral
over vibrational modes,
it is also possible to estimate the ion-thermal free energy from the 
compressibility without explicitly calculating any ion-thermal motion
\cite{Moruzzi1988}, although ion-thermal energies deduced in this way
from the cold compression curve have been found to be inaccurate.

It has recently been shown that results from PIMC and QMD
for elements can be reproduced by all-electron DFT calculations of 
spherically-symmetric wavefunctions about an ion situated in a 
cavity within a uniform charge density `jellium' representing neighboring
atoms \cite{Benedict2014,Driver2017,Swift2018}.
This atom-in-jellium approach \cite{Liberman1979} requires much less
computation than three-dimensional, multi-ion simulations,
and the ion-thermal treatment, based on calculating the Debye temperature
\cite{Liberman1990} has been extended to account for the
decrease in heat capacity from 3 to $\frac 32\kB$ per atom as the ions
cease to be bound by their neighbors at high temperatures \cite{Swift2019}.
The atom-in-jellium model is a gross simplification
of the interaction between neighboring nuclei, and the calculation
of ion-thermal energy is based on a crude average vibration energy
compared with QMD and phonon treatments, but
its use of a Debye-like ion-thermal model captures both the zero-point energy
and the freezing of vibrational modes.
These attributes are likely to be masked by the relative inaccuracy of
atom-in-jellium at low pressures, but they are potentially significant
in cool, compressed states as can be accessed using ramp loading
\cite{Smith2014}.
Despite significant inaccuracies around STP,
atom-in-jellium calculations do indeed seem to perform well for
ramp compression into the terapascal regime as is possible at the 
National Ignition Facility \cite{Fratanduono2019}.
We had previously presumed that this consistency mostly reflected the difficulty
of making experimental measurements accurate enough to show the inaccuracy
of the atom-in-jellium model at these high pressures and internal energies,
but recent QMD studies on states of warm dense matter states
have suggested that ions interact effectively through 
a screened Yukawa potential \cite{Stanton2018},
and thus the atom-in-jellium model may actually be close to accurate
in this regime, as opposed merely to being relatively less inaccurate.

\section{Equation of state calculations}
The principal motivation for this study was to test and improve EOS
for use in high energy density (HED) experiments, and here we pay
closest attention to theoretical calculations valid on terapascal
and electron-volt scales, and higher.
Thus we used the atom-in-jellium model to construct wide-range EOS
in the expectation that they would be relatively inaccurate around STP.
Given the interesting observations that Ru and Rh retain their ambient structure
(hexagonal close packed, hcp, and face-centered cubic, fcc, respectively)
to the highest pressures reported experimentally \cite{Cynn2002,Yusenko2019}
or theoretically \cite{Cazorla2008}, this 
suggests their potential use as pressure calibrants in diffraction experiments.
With the development of ramp loading and {\it in situ} diffraction, 
such experiments are now routinely performed in large laser facilities 
into the terapascal range \cite{rampdiff}, and so
we used multi-ion pseudopotential calculations of the cold compression curve
for several crystal structures as an indication of whether phase transitions
might be expected at pressures above those investigated previously.
We used these cold curves to estimate the Gr\"uneisen parameter
and hence, combined with the experimentally-determined $\thetaD$, 
an estimate of the ion-thermal contribution to the EOS,
providing greater accuracy around STP in the solid.

\subsection{Atom-in-jellium}
Atom-in-jellium simulations were performed
using the same prescription as developed previously for other elements
\cite{Swift2018,Swift2019}.
For each element, atom-in-jellium calculations were made over a range
and density of states suitable for a general-purpose EOS:
mass density $\rho$ from
$10^{-4}$ to $10^3\rho_0$ with 20 points per decade,
and temperature $T$ from $10^{-3}$ to $10^5$\,eV with 10 points per decade.
The reference mass density $\rho_0$ was chosen to be the observed STP value;
this choice is purely a convenience in constructing tabular EOS,
where it is useful for the tabulation
to include the starting state to reduce the sensitivity to interpolating
functions.
The EOS were not adjusted to reproduce any empirical data.

As was found in the previous study \cite{Swift2018}, the 
electronic wavefunctions were computed reliably down to 10\,K or less
for densities corresponding to condensed matter, and to 100\,K or less for densities
down to 0.1\%\ of the ambient solid.
At lower densities, calculations were completed successfully only for 
temperatures of several eV or more.

The results of the atom-in-jellium calculations were, 
for each state of mass density $\rho$ and temperature $T$,
electronic contrbutions to the Helmholtz free energy $f$,
the estimated Debye temperature $\thetaD$, 
the mean square displacement of the atom as a fraction of the Wigner-Seitz
radius $f_d$,
and the ionic contribution to $f$ using the generalized Debye model
with asymptotic ionic freedom \cite{Swift2018,Swift2019}.
The total electronic energy was used: it was taken to include the
cold compression energy and was not adjusted to extract a separate 
electron-thermal energy.
The fields were post-processed to fill in states where the calculation
failed to converge properly,
using polynomial interpolation from surrounding states.
For each state, the total Helmholtz free energy $f$ was calculated, and then 
differentiated using a quadratic fit to the three closest states in $\rho$ to
determine the pressure $p(\rho,T)$ in tabular form.
Similarly, quadratic fits in $T$ were differentiated to find the specific
entropy $s$ and hence the specific internal energy $e(\rho,T)$ in tabular form.
These tabulated functions comprise an EOS in {\sc sesame} or {\sc leos} form.

To help assess the accuracy of the ion-thermal energy, we considered
the predicted Debye temperatures $\thetaD$.
As described previously \cite{Liberman1990,Swift2018,Swift2019},
the calculation proceeds by using perturbation theory to calculate the
force constant for displacement of the ion from the center of the cavity
in the jellium, and hence the Einstein temperature $\thetaE$.
$\thetaD$ is then deduced from $\thetaE$ in two alternative ways,
by equating either the energy or the mean amplitude of displacement.
$\thetaD$ is calculated as a function of both $\rho$ and $T$, although
its precise value is important only when $T\simeq\thetaD$.
For this reason, we have previously calculated an effective $\thetaD(\rho)$
by taking the locus $\thetaD(\rho,T)=T$ \cite{Swift2018}.
The value calculated this way can be significantly different than
the value along the cold compression curve, $\thetaD(\rho,0)$,
or as close to the cold curve as can be reached using atom-in-jellium
computations.

The atom-in-jellium calculations included the mean amplitude $u$ of
ionic oscillations as a function of the Wigner-Seitz radius $\rWS$.
The ratio was used to
predict the variation of melt temperature with compression 
and hence, using the EOS, the melt locus as a function of pressure,
as was reported previously for Al and Fe \cite{Swift_melt_2019}.

\subsection{Multi-atom pseudopotential}
Three-dimensional multi-ion simulations were performed using 
non-local pseudopotentials to represent the inner electrons on each atom,
and a plane wave expansion to represent the outer electrons,
solving the Kohn-Sham DFT equations \cite{Hohenberg64,Kohn65,Perdew92,White94}
with respect to the Schr\"odinger Hamiltonian,
and thus calculating the ground state energy along with forces
and stresses via the Hellmann-Feynman theorem,
for a series of different values of the lattice parameters.

The pseudopotentials used were generated by the Troullier-Martins method
\cite{Troullier91} with the inner 36 electrons (i.e. the closed shells of Kr)
treated as core and the outer 8 (Ru) or 9 (Rh) treated explicitly as valence.
The wavefunction was evaluated at $10^3$ regularly-spaced points in
reciprocal space, reduced by the symmetry of the crystal lattice 
\cite{Monkhorst76}.
A plane-wave cutoff of 2000\,eV was sufficient to converge the ground states
to $\sim$1\,meV/atom or better.

For non-cubic structures,
the lattice vectors giving any specific stress, such as isotropic, 
were predicted by assigning a fictitious mass to the parameters
and evolving them for short periods under the instantaneous Hellmann-Feynman
stress.
The rate of change of each parameter was reset to zero before each 
iteration, so this procedure is effectively a critically-damped dynamics.
The method used was a variant of one developed previously \cite{Warren1997},
modified to check the degree of convergence of the electron states
and stress automatically, rather than performing a set number of iterations.
We found that the number of iterations needed to reach convergence could 
sometimes vary by a factor of several between calculations at adjacent
lattice parameters, highlighting the importance of this adaptive method
in obtaining consistently-converged solutions.

\subsection{Semi-empirical ion-thermal energy}
The multi-ion compression curves were used to estimate the ion-thermal EOS
simply, without having to perform the relatively laborious procedure
of constructing the phonon density of states
as we have in previous work \cite{Swift2001,Swift2007}.
We consider this justified as, in the solid, 
the ion-thermal contribution is a relatively small correction to the 
dominant contribution of the cold curve to the EOS,
and in the present work we are not concerned with subtle effects on
phase boundaries where small corrections may matter.

We used the Burakovsky and Preston's form \cite{Burakovsky2004} 
of the relationship between the Gr\"unseisen parameter $\Gamma(\rho)$
and the cold curve,
\begin{equation}
\Gamma(\rho)=\dfrac{\frac{B'(\rho)}2-\frac 16-\frac t2\left[1-\frac{p(\rho)}{3B(\rho)}\right]}{1-\frac {2t}3\frac{p(\rho)}{B(\rho)}}
\end{equation}
where $p$ is the pressure, $B$ the bulk modulus, and $B'$ its pressure
derivative.
The ion-thermal EOS has been found to be represented most accurately by
a value of $t$ which increases from 0 to 2 with compression
\cite{Burakovsky2004}.
For simplicity, we chose $t=2$ to select the Vashchenko-Zubarev relation
\cite{Vashchenko1963}, as the ion-thermal contribution 
to the principal shock Hugoniot is in any case smaller at lower pressure.

The Gr\"uneisen parameter is the logarithmic derivative of the Debye temperature,
\begin{equation}
\Gamma(\rho)=\dfrac{\rho}{\thetaD}\diffl{\thetaD}\rho,
\end{equation}
and so the Debye temperature can be expressed as
\begin{equation}
\thetaD(\rho)=\thetaD(\rho_r)G(\rho)/G(\rho_r)
\label{eq:coldthetad}
\end{equation}
where
\begin{equation}
G(\rho)\equiv\exp\int^\rho \dfrac{\Gamma(\rho')}{\rho'}d\rho'
\end{equation}
and $\rho_r$ is some reference density.

The electronic structure calculations gave the ground state energy
and, by the Hellmann-Feynman theorem, $p$, at each of a series of 
mass densities $\rho$.
In order to perform the further differentiation necessary to calculate
$B$ and $B'$, the cold curve was fitted with analytic functions.
Several functions were tried, and a modified version of the 
Vinet form \cite{Vinet1987}
was found to fit the cold curve over the widest range:
\begin{equation}
e(\rho)=e_c\phi[a(\rho)]+e_0 : \phi(a)=-\left(1+a+\frac 1{20}a^3\right)
\end{equation}
where 
\begin{equation}
%a(\rho)&=&\dfrac{\left(\rho_0/\rho\right)^{1/3}-1}{l(\rho)} \\
a(\rho)=\dfrac{\left(\rho_0/\rho\right)^{1/3}-1}{l},\quad
%l(\rho)&=&P_n[\eta(\rho)] \\
\eta(\rho)=\dfrac \rho{\rho_0}-1.
\end{equation}
%$P_n$ is a polynomial,
The fitting parameters are $l$, $\rho_0$, $e_c$, and $e_0$.
The modifications to remove the need for material-specific
constants such as the atomic weight, used to calculate the
Wigner-Seitz radius, and thus give a more convenient relation 
for describing arbitrary $e(\rho)$ data, given that the scaling parameter
$l$ is fitted to reproduce the data in any case.
In some cases, such as Rh, the range of the fit can be extended by making
$l$ a function of $\rho$, such as a low-order polynomial.

One advantage of this procedure over a more detailed phonon calculation 
resulting in a tabular EOS
is that it can more readily be adjusted to better match other data.
Adjustments can be made conveniently to the parameters fitting the cold curve,
and to the reference value of the Debye temperature, $\thetaD(\rho_r)$.

When used over the full range of states considered here,
we generally had to split the fit into low and high pressure regions,
with a different set of parameters for each.

\section{Results}
Atom-in-jellium EOS, and multi-atom pseudopotential cold curves,
were constructed for each element as above.
In each case, the EOS was analyzed to extract the ambient isotherm and isochore,
and to deduce the principal isentrope and principal shock Hugoniot,
for comparison with previous EOS and experimental measurements.
These loci are characteristic of different conditions to which matter may be
subjected:
the isotherm and isentrope are limiting cases of compression at low rates,
the isochore is representative of ablation, and the Hugoniot is the locus
for high loading rates.
On the scale of the graphs below, the principal isentrope was not
distinguishable from the ambient isotherm, and is not shown.

%The Rankine-Hugoniot equations describe the conservation of mass,
%momentum, and energy across a steady shock wave:
%\begin{eqnarray}
%\rho_0 u_s & = & \rho(u_s-u_p), \\
%p_0+\rho_0 u_s^2 & = & p+\rho(u_s-u_p)^2, \\
%\left[p_0+\rho_0e_0+\frac 12\rho_0 u_s^2\right]u_s & = & 
%\left[p+\rho e+\frac 12\rho (u_s-u_p)^2\right](u_s-u_p)
%\end{eqnarray}
%where $u_s$ is the shock speed and $u_p$ the particle speed.
%They are typically used to deduce the Hugoniot, or locus of states
%accessible from a given initial state by the passage of a single, steady
%shock wave, for matter described by an EOS of the form $p(\rho,e)$.
%In common with any EOS defined implicitly via a Helmholtz free energy,
%for which the independent variables are $\rho$ and $T$ rather than $\rho$ and
%$e$, for any given specific internal energy $e$, the temperature $T$ is 
%eliminated when solving the Rankine-Hugoniot equations \cite{rhnew}.

To accommodate the wide ranges calculated, we plot graphs in log space.
This turns out to be reasonable:
lines are almost coincident when quantities match to within a
few percent, which is typical for what constitutes agreement in high pressure
physics, while being convenient for quantifying the difference between
models that differ significantly.

\subsection{Rhodium}
\begin{figure}
% Rh/inferno/81j/zstar1a.eps
\begin{center}\includegraphics[scale=0.72]{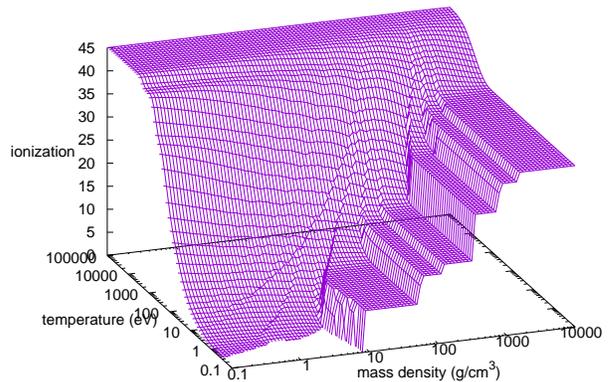}\end{center}
\caption{Atom-in-jellium prediction of ionization in rhodium.}
\label{fig:rhzstar}
\end{figure}

The thermal ionization calculated using the atom-in-jellium method was 
broad with smeared-out shell features, whereas
ionization in response to compression 
(commonly, if loosely, referred to as `pressure ionization') 
was calculated to occur in distinct steps,
a characteristic behavior of the average atom treatment
(Fig.~\ref{fig:rhzstar}).
However, the pressure ionization steps did not manifest as
an abrupt change of slope in the ambient isotherm.
Ionization can be used as an indication of when pseudopotentials 
may become invalid.
In the atom-in-jellium calculations, the Kr-like shells remained bound 
up to 50\,g/cm$^3$ and below 7\,eV.

\begin{figure}
% Rh/inferno/81j/isothdplo.eps
\begin{center}\includegraphics[scale=0.72]{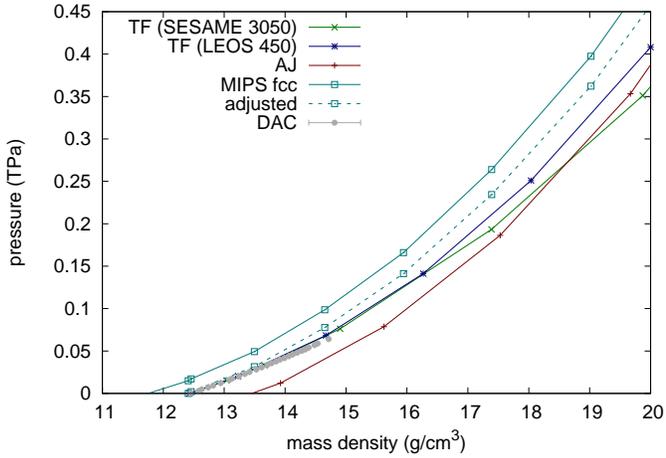}\end{center}
\caption{Low pressure ambient isotherm of rhodium,
   showing atom-in-jellium (AJ) and Thomas-Fermi (TF) based EOS,
   multi-ion pseudopotential calculations (MIPS), diamond anvil cell (DAC)
   measurements \cite{Yusenko2019},
   and pseudopotential calculations adjusted to match the mass density observed
   at STP.}
\label{fig:rhisothdplo}
\end{figure}

At the observed STP mass density of 12.41\,g/cm$^3$, 
the atom-in-jellium EOS gave a pressure of -27\,GPa
and the multi-ion pseudopotential calculation gave 15\,GPa.
These discrepancies are typical for the respective techniques:
somewhat larger than usual for pseudopotential calculations,
and less than usual for atom-in-jellium.
The ambient isotherms lay correspondingly below and above reported 
diffraction data from diamond anvil compression \cite{Yusenko2019}.
To obtain a more accurate EOS at low pressures, we have previously
adjusted the energy obtained from electronic structure calculations
to reproduce the observed STP state \cite{Swift2001,Swift_C_2019}.
Applying this correction to the pseudopotential states brought
the isotherm into better agreement with the data, though the
calculations lay increasingly above the data with pressure.
Isotherms from the general-purpose TF-based EOS deviated above $\sim$0.15\,TPa.
Measurements at higher pressure would be valuable.
(Fig.~\ref{fig:rhisothdplo}.)

\begin{figure}
% Rh/inferno/81j/isothdp1.eps
\begin{center}\includegraphics[scale=0.72]{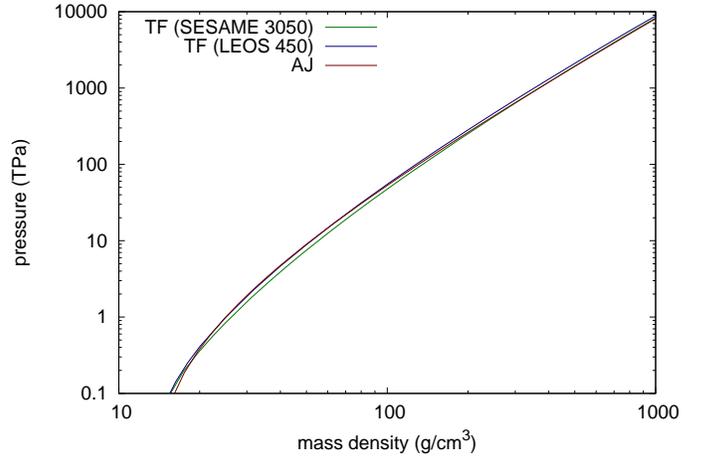}\end{center}
\caption{Ambient isotherm of rhodium from atom-in-jellium (AJ) and Thomas-Fermi (TF) based EOS.}
\label{fig:rhisothdp}
\end{figure}

Over a wider pressure range, 
the isotherm predicted from atom-in-jellium was very consistent 
with the semi-empirical TF-based EOS {\sc leos} 450; 
the similarly-constructed {\sc sesame} 3050 was up to 15\%\ softer.
(Fig.~\ref{fig:rhisothdp}.)

\begin{figure}
% Rh/inferno/81j/isochtp1.eps
\begin{center}\includegraphics[scale=0.72]{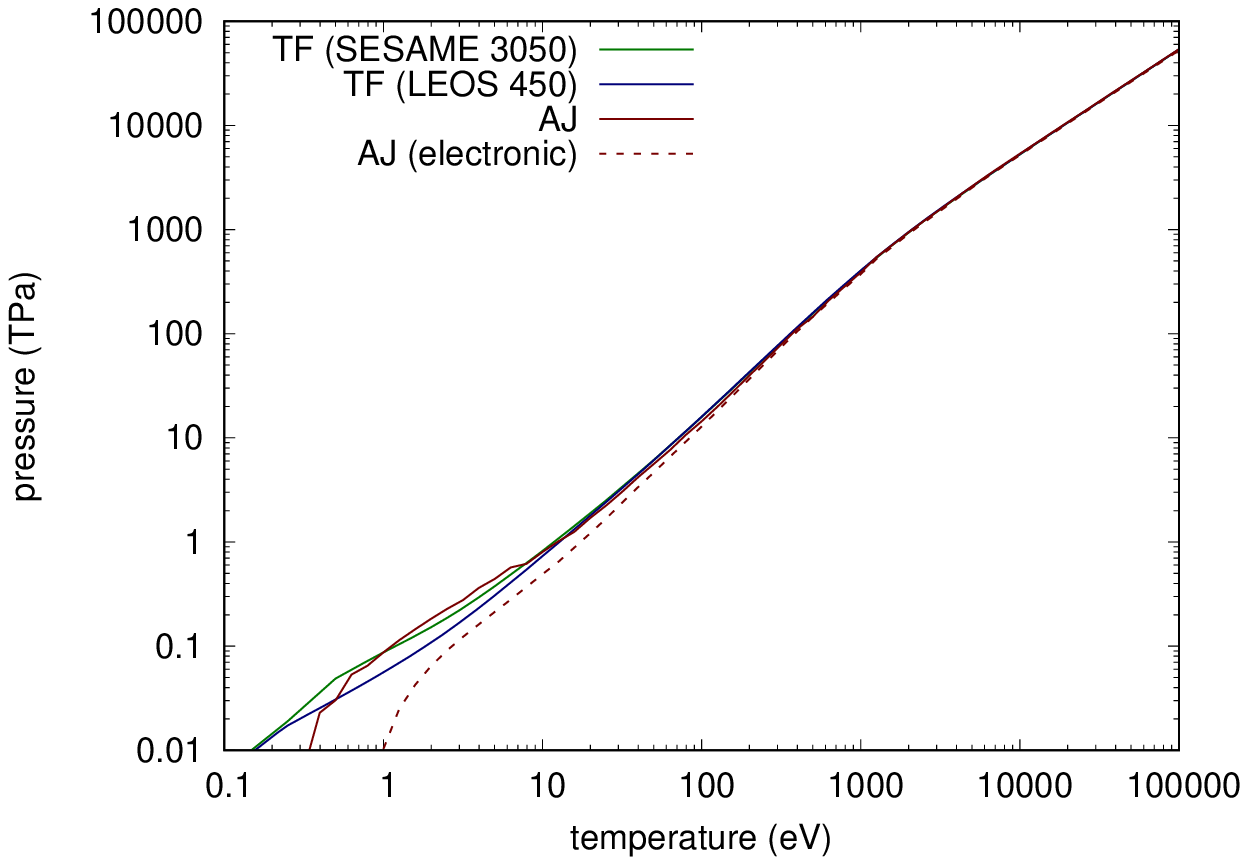}\end{center}
\caption{Ambient isochore of rhodium from atom-in-jellium (AJ) and Thomas-Fermi (TF) based EOS.}
\label{fig:rhisochtp}
\end{figure}

The atom-in-jellium isochore closer to the {\sc sesame} EOS up to $\sim$10\,eV, 
though different from both and showing ionization of successive electron shells.
The atom-in-jellium EOS should be as reasonable as either of the
semi-empirical EOS to below 30\,GPa, 
when ionic contribution dominates.
This is a remarkable degree of consistency
of ion-thermal pressure from the atom-in-jellium model.
At higher temperatures relevant to ablation in laser-driven experiments,
all the EOS were close to each other.
(Fig.~\ref{fig:rhisochtp}.)

\begin{figure}
% Rh/inferno/81j/hugdp1.eps
\begin{center}\includegraphics[scale=0.72]{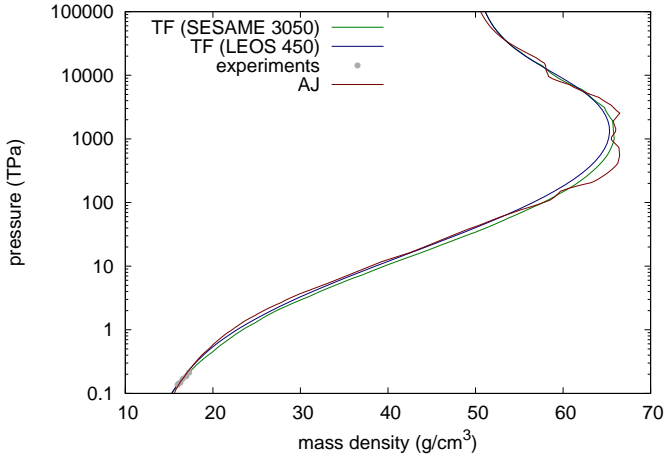}\end{center}
\caption{Principal Hugoniot of rhodium from atom-in-jellium (AJ) and Thomas-Fermi (TF) based EOS.}
\label{fig:rhhugdp}
\end{figure}

The atom-in-jellium Hugoniot passed among the experimental measurements 
around 0.2\,TPa.
Below 0.1\,TPa, it fell well below the experiments, reflecting the
inaccuracy of atom-in-jellium around STP.
The atom-in-jellium Hugoniot was $\sim$10\%\ stiffer than the 
semi-empirical {\sc leos}~450 at a few terapascals,
remarkably close to it over $\sim$8-70\,TPa, and then exhibited
shell ionization features at higher pressures, not predicted with TF models.
The other TF-based semi-empirical EOS, 
{\sc sesame}~3050, generally lay $\sim$10-20\%\ below
the atom-in-jellium and {\sc leos} Hugoniots.
(Fig.~\ref{fig:rhhugdp}.)

The pseudopotential calculations predicted that the bcc structure
was energetically unfavorable over the full compression range
considered, but that fcc and hcp lay much closer together.
Fitting a modified Vinet function to the fcc calculations,
which was valid to at least 45\,g/cm$^3$,
the hcp structure was found to lie at a higher energy even when $c/a$
was optimized, and the calculations were thus consistent with Rh
remaining in the fcc structure to at least 10\,TPa, which is the limit
beyond which the pseudopotential was likely to become invalid.
This result extends the previous study using tight binding theory
\cite{Cazorla2008} by a factor $\sim$20 in pressure.

\begin{figure}
% Rh/inferno/81j/dispm.eps
\begin{center}\includegraphics[scale=0.72]{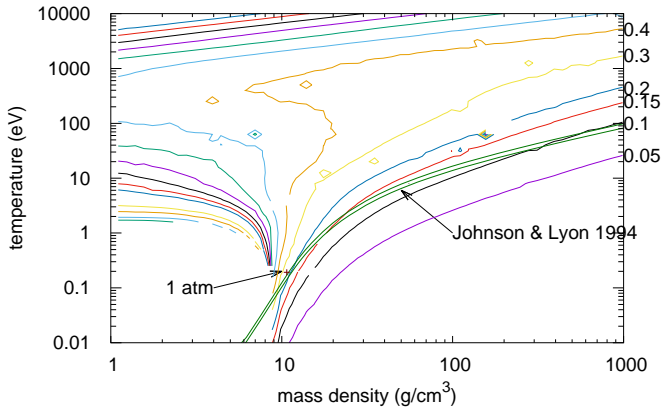}\end{center}
\caption{Contours of mean ionic displacement of Rh divided by the Wigner-Seitz radius,
   along with the observed one-atmosphere melt point and a modified Lindemann
   prediction of the melt locus (solidus and liquidus) \cite{Johnson1994a}.}
\label{fig:rhdispm}
\end{figure}

\begin{figure}
% Rh/inferno/81j/meltpt.eps
\begin{center}\includegraphics[scale=0.72]{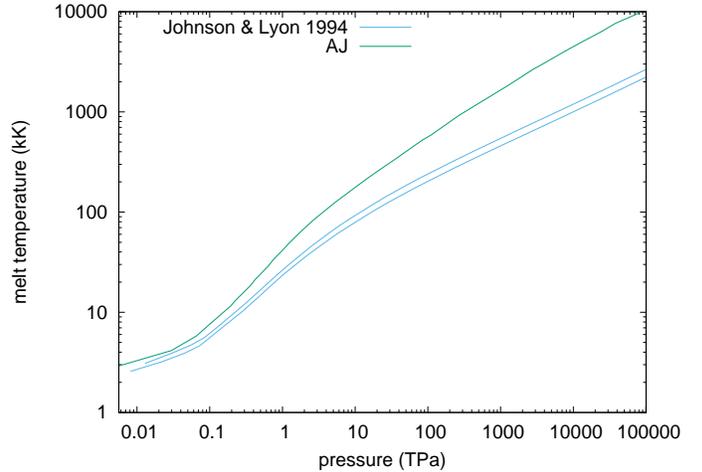}\end{center}
\caption{Melt locus for Rh predicted from atom-in-jellium oscillations (AJ)
   and using a modified Lindemann model \cite{Johnson1994a}.}
\label{fig:rhmeltpt}
\end{figure}

One-atmosphere melting of Rh lay close to the ion oscillation contour
with an amplitude of $u/\rWS\simeq 0.2$.
Taking this contour to extrapolate to other compressions,
we predicted the melt locus as a function of mass density and, using the EOS,
a function of pressure.
The melt locus of Rh has been predicted previously using a modified Lindemann
criterion in which the Gr\"uneisen parameter is used to extrapolate
from the one-atmosphere melt temperature by integration
\cite{Johnson1994a}.
The atom-in-jellium prediction lay significantly above the older model.
(Figs~\ref{fig:rhdispm} and \ref{fig:rhmeltpt}.)

\subsection{Ruthenium}
\begin{figure}
% Ru/inferno/81j/zstar1a.eps
\begin{center}\includegraphics[scale=0.72]{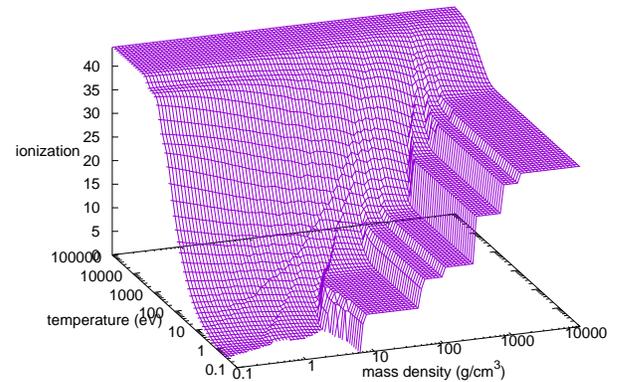}\end{center}
\caption{Atom-in-jellium prediction of ionization in ruthenium.}
\label{fig:ruzstar}
\end{figure}

Atom-in-jellium predictions of ionization in Ru exhibited 
features very similar to those in Rh.
The Kr-like shells remained bound to above 40\,g/cm$^3$ and below 10\,eV
(Fig.~\ref{fig:ruzstar}).

\begin{figure}
% Ru/inferno/81j/isothdplo.eps
\begin{center}\includegraphics[scale=0.72]{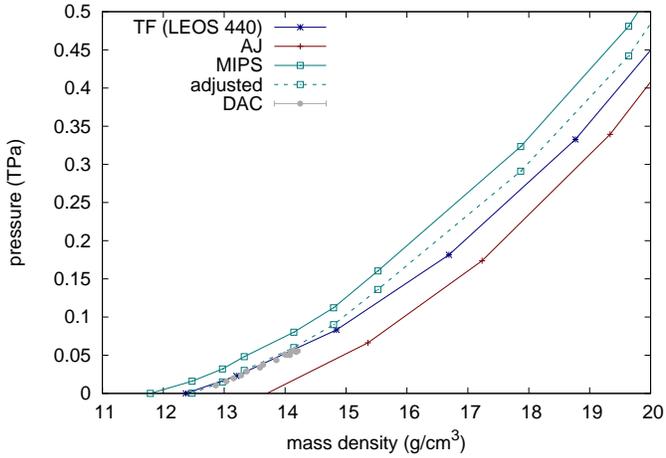}\end{center}
\caption{Low pressure ambient isotherm of ruthenium,
   showing atom-in-jellium (AJ) and Thomas-Fermi (TF) based EOS,
   multi-ion pseudopotential calculations (MIPS), diamond anvil cell (DAC)
   measurements \cite{Cynn2002},
   and pseudopotential calculations adjusted to match the mass density observed
   at STP.}
\label{fig:ruisothdplo}
\end{figure}

At the observed STP mass density of 12.45\,g/cm$^3$, 
the atom-in-jellium EOS gave a pressure of -39\,GPa
and the multi-ion pseudopotential calculation gave 16\,GPa:
similar discrepancies as for Rh.
The ambient isotherms lay correspondingly below and above reported 
diffraction data from diamond anvil compression \cite{Cynn2002}.
To obtain a more accurate EOS at low pressures, we have previously
adjusted the energy obtained from electronic structure calculations
to reproduce the observed STP state \cite{Swift2001,Swift_C_2019}.
Applying the equivalent correction \cite{Swift2001,Swift_C_2019}
as for Rh to bring the pseudopotential calculations into agreement with the
STP state, the adjusted isotherm matched the diamond anvil data well.
Interestingly, the isotherm deviated from the 
TF-based {\sc leos}~440, the only semi-empirical EOS available,
starting just above the limit of diamond anvil data;
measurements to higher pressure would be valuable.
(Fig.~\ref{fig:ruisothdplo}.)

\begin{figure}
% Ru/inferno/81j/isothdp1.eps
\begin{center}\includegraphics[scale=0.72]{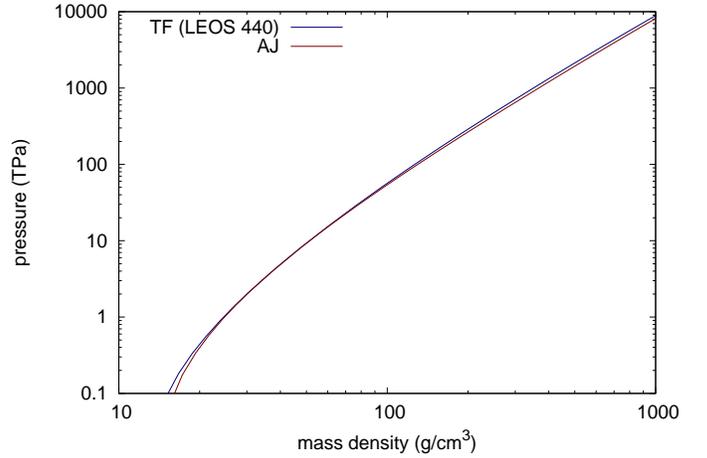}\end{center}
\caption{Ambient isotherm of ruthenium from atom-in-jellium (AJ) and Thomas-Fermi (TF) based EOS.}
\label{fig:ruisothdp}
\end{figure}

\begin{figure}
% Ru/inferno/81j/ambisoch.eps
\begin{center}\includegraphics[scale=0.72]{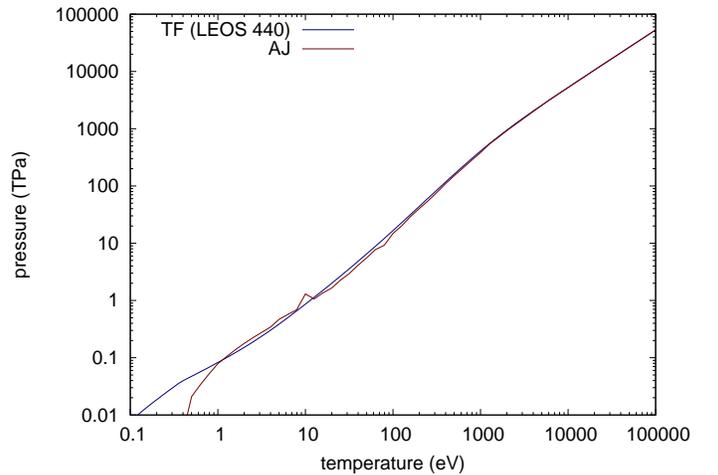}\end{center}
\caption{Ambient isochore of ruthenium from atom-in-jellium (AJ) and Thomas-Fermi (TF) based EOS.}
\label{fig:ruisochtp}
\end{figure}

Over a wider pressure range,
the ambient isotherm and isochore from atom-in-jellium
were remarkably similar to that from the TF-based {\sc leos}~440, 
though not identical.
The agreement between the isotherms is notable as, at low pressures,
semi-empirical EOS are dominated by fitting to shock data, 
which was non-existent for Ru.
(Figs~\ref{fig:ruisothdp} and \ref{fig:ruisochtp}.)

\begin{figure}
% Ru/inferno/81j/hugdp1.eps
\begin{center}\includegraphics[scale=0.72]{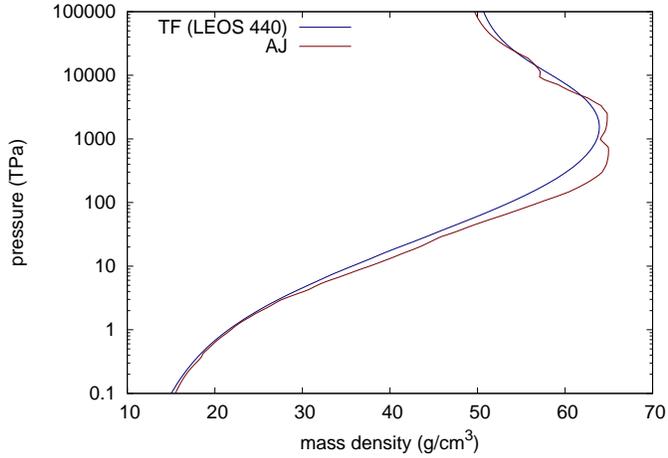}\end{center}
\caption{Principal Hugoniot of ruthenium from atom-in-jellium (AJ) and Thomas-Fermi (TF) based EOS.}
\label{fig:ruhugdp}
\end{figure}

Similarly for the principal Hugoniot, the results from atom-in-jellium
were very similar to the semi-empirical EOS,
indicating that extrapolation from ambient using the bulk modulus 
was surprisingly accurate.
Counterintuitively, given the agreement to $\sim$70\,TPa in Rh,
for Ru the atom-in-jellium Hugoniot fell below the TF model around 3\,TPa, 
the deviation increasing to $\sim$10\%\ in mass density and $\sim$20\%\ in
pressure, before the pronounced effects of shell ionization gave larger
differences above 100\,TPa.
(Fig.~\ref{fig:ruhugdp}.)

\begin{figure}
% Ru/castep/hcpc_rel/cacmp.eps
\begin{center}\includegraphics[scale=0.72]{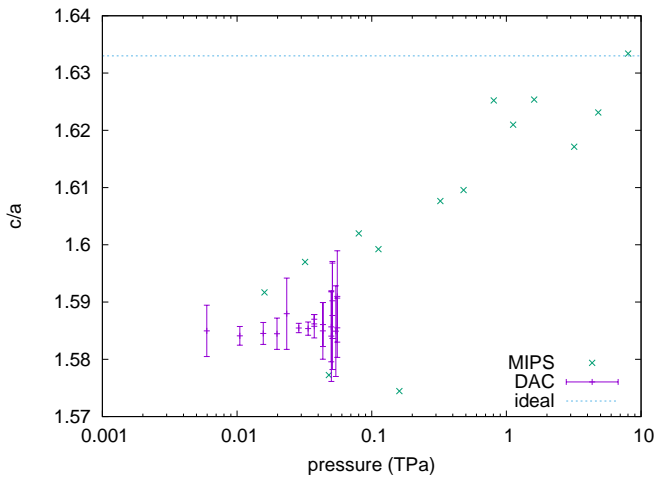}\end{center}
\caption{$c/a$ for hcp ruthenium, from multi-ion pseudopotential (MIPS)
   calculations, 
   compared with diamond anvil cell (DAC) measurements \cite{Cynn2002}.}
\label{fig:ruca}
\end{figure}

Using the three-dimensional, multi-ion pseudopotential method,
the lattice parameters for states of isotropic stress were calculated 
as described above.
Lattice vectors were sought for a sequence of increasing pressures,
using the previous lattice vectors as the initial guess for the next higher
pressure.
Converged to finite accuracy, the lattice parameters exhibited clear
trends, but also showed numerical noise.
The hcp $c/a$ ratio was calculated to be 1.583 at zero pressure,
close to the observed value of 1.584 \cite{Cynn2002}
and to the value 1.58 found
using full-potential linearized augmented plane wave (FLAPW) calculations
\cite{Cazorla2008}.
$c/a$ was predicted to rise to the ideal value of $\sqrt{8/3}$ around 10\,TPa, 
and then to fall for higher pressures. 
However, above 10\,TPa, the density reaches the range where 
the atom-in-jellium calculations predicted
pressure-ionization of the $4p$ shell,
and so a deeper pseudopotential should be used.
The rate of increase of $c/a$ with pressure was locally faster than
observed in diamond anvil cell experiments,
but had a similar magnitude \cite{Cynn2002}.
(Fig.~\ref{fig:ruca}.)

The pseudopotential calculations predicted that hcp remained
the lowest energy structure of those considered,
up to the highest compression calculated of 87.5\,g/cm$^3$ or 80.1\,TPa.
This result was obtained allowing the hcp structure to relax, i.e. $c/a$
to vary;
with $c/a$ held constant at the observed STP value, fcc was calculated to
become more stable between 0.42 and 2.26\,TPa.
Above 2.5\,TPa, the energies of fcc and the relaxed hcp structures were very
close, so the prediction of hcp stability may be unreliable,
and above 10\,TPa the prediction should be regarded as unreliable
since the calculations did not account for ionization of the inner shells.
The bcc structure was calculated to have a significantly higher energy than
hcp or fcc over full range considered, consistent with the FLAPW results
at lower pressures \cite{Cazorla2008}.

\begin{figure}
% Ru/inferno/81j/thetadcmp.eps
\begin{center}\includegraphics[scale=0.72]{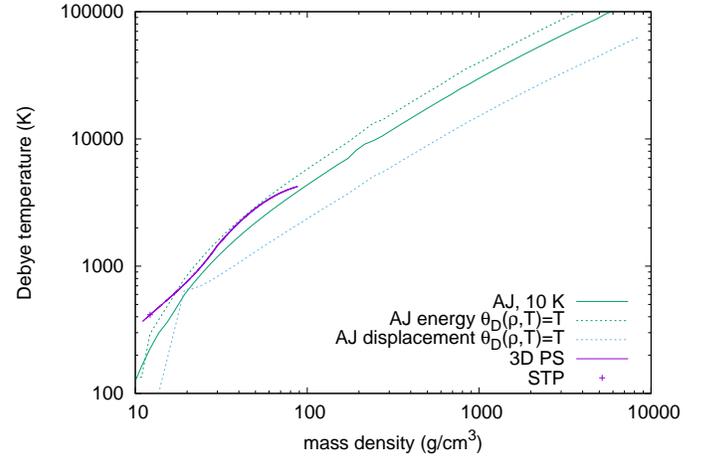}\end{center}
\caption{Debye temperature of ruthenium, from atom-in-jellium calculations
   and also inferred from the STP value using the Gr\"uneisen $\Gamma$
   calculated from the 3D pseudopotential calculations.}
\label{fig:ruthetad}
\end{figure}

Debye temperatures $\thetaD$ were estimated from the hcp cold curve,
using Eq.~\ref{eq:coldthetad}
to predict the variation from the STP value of 415\,K \cite{Ho1974},
and compared with the alternative atom-in-jellium prescriptions for $\thetaD$.
The atom-in-jellium $\thetaD(\rho,10\,\mbox{K})$, effectively the cold curve
dependence, was indistinguishable using the energy and displacement
prescriptions.
The loci calculated as $\thetaD(\rho,T)=T$ in each case, which should be
more accurate as a representation of the full $\thetaD(\rho,T)$,
varied from the cold curve dependence in opposite directions.
Apart from the region below 20\,g/cm$^3$, where the atom-in-jellium $\thetaD$
fell well below, the $\thetaD$ estimated from the multi-ion simulations
lay between the cold-curve and energy-based effective $\thetaD$ model,
which was higher than the cold variation, and closer to the effective
$\thetaD$.
At higher compressions, where the pseudopotential was expected to become
invalid, $\thetaD$ was predicted to flatten out;
this result is probably not reliable.
This comparison suggests that $\thetaD$ calculated in these very different
ways can be consistent, at least at the few percent level,
and that the energy-based atom-in-jellium calculation of $\thetaD$ may be
the better.
(Fig.~\ref{fig:ruthetad}.)

\begin{figure}
% Ru/inferno/81j/disp1.eps
\begin{center}\includegraphics[scale=0.72]{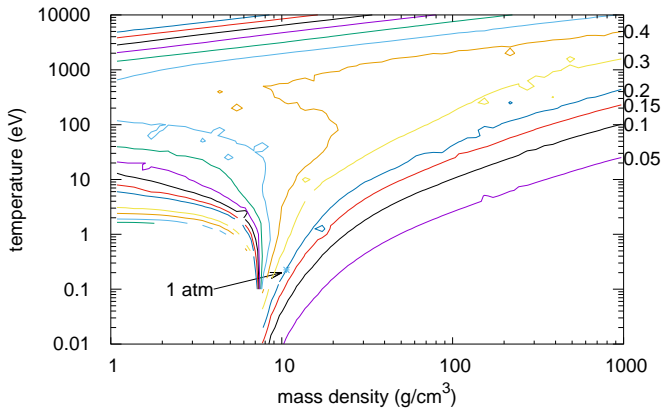}\end{center}
\caption{Contours of mean ionic displacement of Ru divided by the Wigner-Seitz radius,
   along with the observed one-atmosphere melt point.}
\label{fig:rudispm}
\end{figure}

\begin{figure}
% Ru/inferno/81j/meltpt.eps
\begin{center}\includegraphics[scale=0.72]{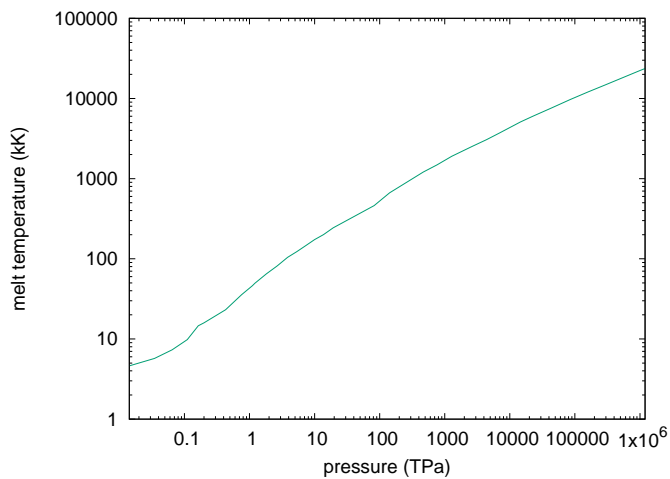}\end{center}
\caption{Melt locus for Ru predicted from jellium oscillations.}
\label{fig:rumeltpt}
\end{figure}

One-atmosphere melting of Ru also lay close to the ion oscillation contour
with an amplitude of $u/\rWS\simeq 0.2$.
As for Rh, we used this contour to extrapolate to other pressures.
We attempted to fit the Simon relation \cite{Simon1929} to the locus;
it was not able to fit the full range of the calculation, but was a reasonable
fit in each of three regions demarcated by 10 and 1000\,TPa.
Interestingly, these points correspond to the most abrupt changes in
pressure ionization, suggesting that the melt locus appears to be more directly
sensitive to changes in ionization than the isotherm.
(Figs~\ref{fig:rudispm} and \ref{fig:rumeltpt}.)

\section{Conclusions}
Wide-ranging EOS, intended principally for high energy density applications,
were constructed self-consistently for Ru and Rh using atom-in-jellium theory.
In Rh, for which measurements of the shock Hugoniot have been reported
and used previously in the construction of general-purpose EOS,
our {\it ab initio} EOS gave very similar behavior,
except that our Hugoniot exhibited features corresponding to the
ionization of successive electron shells which were absent in the
previous EOS based on TF theory.
In Ru, for which no shock Hugoniot measurements have been reported,
our EOS was significantly more compressible than the only 
general-purpose EOS known,
suggesting that this EOS gives shock pressures that are roughly 40\%\ too high,
or mass densities that are about 10\%\ too low, in the range $\sim$4-500\,TPa.
Interestingly, in the regime where the semi-empirical EOS would have
been guided by shock data, it agreed very well with the {\it ab initio} EOS;
the difference arose at higher shock pressures where the semi-empirical
EOS transitioned to the TF model.
These comparisons indicate that the QEOS procedure used to 
construct the semi-empirical EOS in general works extremely well.
Nevertheless, this difference in EOS between two adjacent transition elements 
highlights the potential inaccuracy of semi-empirical EOS 
constructed given limited shock data, or in its absence,
and is a motivation to collect additional data 
in the terapascal range and above.

Cold compression curves were predicted for the common crystal structures 
using three dimensional pseudopotential calculations.
Ru was predicted to remain hcp to at least 2.5\,TPa, probably 10\,TPa, 
and possibly higher,
although the pseudopotentials used to calculate these cold curves
are likely to be inaccurate at higher pressures.
$c/a$ matched the observed value at STP, and was predicted to increase
toward the ideal value at $\sim$10\,TPa;
the initial rate of increase matched existing diamond anvil results.
Rh was predicted to remain fcc to at least 10\,TPa.
The predicted absence of phase transitions in both elements suggests 
their possible use
as a pressure calibrant in high-pressure diffraction experiments.

Melt loci were predicted for both elements;
the result for Rh lay significantly above the previous prediction.

\section*{Acknowledgments}
This work was performed under the auspices of
the U.S. Department of Energy under contract DE-AC52-07NA27344.

\end{document}